\documentclass[runningheads]{llncs}
\usepackage{graphicx}
\usepackage{amsmath}
\usepackage{siunitx}
\usepackage{gensymb}
\usepackage[inline]{fixme}
\usepackage{stmaryrd}
\usepackage[T1]{fontenc}
\usepackage{hyperref}

\begin{document}
\title{Slice estimation in diffusion MRI of neonatal and fetal brains in image and spherical harmonics domains using autoencoders}
\titlerunning{dMRI neonatal and fetal brain slice estimation}

\authorrunning{H. Kebiri et al.}
%
\author{Hamza Kebiri \inst{1,2} \and Gabriel Girard\inst{1,2,3} \and
Yasser Alem\'an-G\'omez\inst{1} \and Thomas Yu\inst{3,4} \and Andr\'as Jakab\inst{5,6} \and Erick Jorge Canales-Rodr\'iguez\inst{3}  \and Meritxell Bach Cuadra\inst{2,1}}

\institute{********\\******}

\institute{Department of Radiology, Lausanne University Hospital and University of Lausanne, Lausanne, Switzerland \and CIBM Center for Biomedical Imaging, Switzerland \and Signal Processing Laboratory 5 (LTS5), \'Ecole Polytechnique F\'ed\'erale de Lausanne (EPFL), Lausanne, Switzerland \and Advanced Clinical Imaging Technology, Siemens Healthineers International AG, Lausanne, Switzerland \and Center for MR Research, University Children's Hospital Zurich, Zurich, Switzerland \and Neuroscience Center Zurich, University of Zurich, Zurich, Switzerland}

\maketitle              
\begin{abstract} 
Diffusion MRI (dMRI) of the developing brain can provide valuable insights into the white matter development. However, slice thickness in fetal dMRI is typically high (i.e., 3-5 mm) to freeze the in-plane motion, which reduces the sensitivity of the dMRI signal to the underlying anatomy. In this study, we aim at overcoming this problem by using autoencoders\index{autoencoder} to learn unsupervised\index{unsupervised learning} efficient representations of brain slices in a latent space, using raw dMRI signals and their spherical harmonics\index{spherical harmonics} (SH) representation. We first learn and quantitatively validate the autoencoders\index{autoencoder} on the developing Human Connectome Project pre-term newborn data, and further test the method on fetal data. Our results show that the autoencoder\index{autoencoder} in the signal domain better synthesized the raw signal. Interestingly, the fractional anisotropy\index{fractional anisotropy} and, to a lesser extent, the mean diffusivity\index{mean diffusivity}, are best recovered in missing slices by using the autoencoder\index{autoencoder} trained with SH \index{spherical harmonics} coefficients. A comparison was performed with the same maps reconstructed using an autoencoder\index{autoencoder} trained with raw signals, as well as conventional interpolation\index{interpolation} methods of raw signals and SH\index{spherical harmonics} coefficients. From these results, we conclude that the recovery of missing/corrupted slices should be performed in the signal domain if the raw signal is aimed to be recovered, and in the SH\index{spherical harmonics} domain if diffusion tensor \index{diffusion tensor imaging} properties (i.e., fractional anisotropy\index{fractional anisotropy}) are targeted. Notably, the trained autoencoders\index{autoencoder} were able to generalize to fetal dMRI data acquired using a much smaller number of diffusion gradients and a lower b-value, where we qualitatively show the consistency of the estimated diffusion tensor\index{diffusion tensor imaging} maps.


\keywords{Super-resolution\and Autoencoders\and Spherical Harmonics \index{spherical harmonics} \and Diffusion Tensor Imaging\index{diffusion tensor imaging}\and Pre-term\and Fetal\and Brain \and MRI}
\end{abstract}
\section{Introduction}
Neonatal and fetal brain development involves complex cerebral growth and maturation both for gray and white matter \cite{batalle2018annual,dubois2021mri}. Diffusion MRI (dMRI) has been widely employed to study this developmental process \emph{in vivo}, including neonates and fetuses \cite{huppi2006diffusion,jakab2015disrupted,ouyang2019delineation}. As the diffusion weighted signal is sensitive to the displacement of water molecules, several models have been proposed for estimating the underlying anatomy such as diffusion tensor imaging\index{diffusion tensor imaging} (DTI) or spherical deconvolution methods \cite{basser1994mr,tournier2008resolving,Canales-Rodriguez2019}. The accuracy of these models is dependant on the angular and spatial resolution of the acquisitions that is typically limited for the neonate and fetal subjects \cite{jakab2017utero,kimpton2021diffusion}. Stochastic motion and low signal-to-noise ratio (SNR) due to the small size of the developing brain often translate to degraded images with low spatial resolution. Additionally, slice thickness in fetal dMRI is typically high, varying between 3-5 mm, to freeze the in-plane motion, and hence reduces the sensitivity of the dMRI signal to the underlying anatomy. This highlights the need for methods to interpolate\index{interpolation} or synthesize new slices that were either (1) corrupted because of motion or (2) acquired using anisotropic voxel sizes. Interpolation\index{interpolation} is often performed either at scanner level or in post-processing \cite{jakab2017utero}, and has been demonstrated to be relevant for raw signal recovery and for subsequent analysis such as tractography \cite{dyrby2014interpolation}. Similarly, super-resolution (SR) methods that aim at increasing dMRI resolution can be applied at the acquisition-reconstruction level \cite{ning2016joint,ramos2020high} or at post-processing \cite{blumberg2018deeper,elsaid2019super,chatterjee2021shuffleunet}. The latter used \emph{supervised learning} methods, which require high resolution training data that is often unavailable for the developing brain. Additionally, these methods focus on enhancing the resolution homogeneously over all dimensions and were not assessed for anisotropic voxels, commonly acquired for fetuses and neonates \cite{jakab2017utero,kimpton2021diffusion}. Additionally to the raw dMRI signal interpolation\index{interpolation}, other representations such as Spherical Harmonics\index{spherical harmonics} (SH) could be of interest. SH\index{spherical harmonics} are a combination of smooth orthogonal basis functions defined on the surface of a sphere able to represent spherical signals, such as the dMRI signal acquired using uniformly distributed gradient directions \cite{frank2002characterization,hess2006q}. Previous work used deep learning methods to map the SH\index{spherical harmonics} coefficients from one shell to another \cite{koppers2016diffusion,jha2020multi}. However, no prior work, to the best of our knowledge, relies on the SH\index{spherical harmonics} decomposition to enhance the spatial image resolution.

In this study, we have used \emph{unsupervised learning}\index{unsupervised learning} to extend the application of autoencoders \index{autoencoder} for through-plane super-resolution \cite{sander2021unsupervised,kebiri2021through} in the image domain to spherical harmonics \index{spherical harmonics} domain where we synthesize SH\index{spherical harmonics} coefficients of \emph{missing} slices. As such, our network has access to both angular and spatial information. In contrast to training with non-DWI volumes \cite{kebiri2021through}, we have additionally trained a second network on spherical averaged dMRI images to complement and compare its performance in relation to the SH \index{spherical harmonics} trained network. Moreover, we have compared both methods to conventional interpolation\index{interpolation} methods both using raw dMRI signals and their SH\index{spherical harmonics} representation. The comparison was performed both on the raw dMRI signal; and on fractional anisotropy\index{fractional anisotropy} (FA) and mean diffusivity\index{mean diffusivity} (MD) maps derived from the estimated diffusion tensors\index{diffusion tensor imaging}. Finally, we verified that the SH\index{spherical harmonics} networks trained on pre-term data successfully generalized to fetal images, where we present the coherence of the synthesized slices.

\section{Methodology}
\subsection{Materials}
\setcounter{footnote}{0} 
\textbf{Neonatal data -} The developing Human Connectome Project (dHCP) data\footnote{\href{http://www.developingconnectome.org/data-release/second-data-release/}{http://www.developingconnectome.org/data-release/second-data-release/}} were acquired in a 3T Philips Achieva scanner in a multi-shell scheme (b $\in \{0, 400,1000,2600\}$ s/mm$^{2}$). Details on acquisition parameters can be found in \cite{hutter2018time}. The data was denoised, motion and distortion corrected \cite{bastiani2019automated} and has a final resolution of $1.17 \times 1.17 \times 1.5$ mm$^3$ in a FOV of $128 \times 128 \times 64$ mm$^3$. In addition to $b=0$ s/mm$^{2}$ images (b0), we have selected the corresponding 88 volumes with $b=1000$ s/mm$^{2}$ (b1000) from all pre-term subjects (31) defined with less than 37 gestational weeks (GW) ([29.3, 37.0], mean=35.5). In the anatomical dataset, brain tissue labels and masks \cite{makropoulos2014automatic} were provided.

\textbf{Fetal data -}
The fetal data were acquired with the approval of the ethics committee. Acquisitions were performed at 1.5T (GE Healthcare) with a single shot echo planar imaging sequence (TE=63 ms, TR=2200 ms) using $b=700$ s/mm$^{2}$ (b700) and 15 directions. The acquisition FOV was $256\times256\times14-22$ mm$^3$ for a resolution of $1\times1\times4-5$ mm$^3$. Three axial and one coronal acquisitions were performed for each subject. Four subjects were used in our study: two of 35 and 29 GW where three axial volumes were used, and two young subjects of 24 GW where one axial volume was used. We have only used axial acquisitions to avoid any confounding factor due to interpolation\index{interpolation} in the registration that would be needed between the orthogonal orientations. Volumes were corrected for noise \cite{veraart2016diffusion}, bias-field inhomogeneities \cite{tustison2010n4itk} and distortions \cite{kuklisova2017distortion,avants2009advanced} and did not require any motion correction.
\subsection{Model}
\textbf{Network architecture -} Our network is composed of four blocks in the encoder and four blocks in the decoder, where each block consists of two layers of $3\times3$ convolutions, a batch normalization and an Exponential Linear Unit (ELU) activation function \cite{clevert2015fast}. After each block of the encoder, a $2\times2$ average pooling operation was performed and the number of feature maps was doubled after each layer. Hence starting from 32 feature maps to 256 while three additional 3x3 convolutions were added in the last block with 512, 256 and M feature maps respectively, $M \in \{16,32,64,128\}$. The last $M$ feature maps were considered as the latent space of our autoencoder\index{autoencoder}. The decoder goes back to original input dimensions by means of either $3\times3$ transposed convolutions with strides of 2 or by $2\times2$ nearest neighbor interpolations\index{interpolation} (mutually exclusive), where the number of feature maps decreases by two after each layer from 512 to 32. A last 1x1 convolution with sigmoid activation function was performed to generate the predicted image.

\textbf{Training -} Using the same architecture, we have trained three networks, with different inputs: b0 images (\emph{b0-net}), average b1000 (\emph{Avg-b1000-net}) (see \emph{Raw signal networks} subsection) and a maximum SH\index{spherical harmonics} order ($L_{max}$) of 4 (\emph{SH4-net}) (see \emph{Spherical harmonics networks}\index{spherical harmonics} subsection). Input images were first normalized to the range $[0,1]$ by $x = \frac{x - x_{min}}{x_{max} - x_{min}}$ where $x_{min}$ and $x_{max}$ are the minimum and maximum intensities respectively in a given slice. All networks were trained using an Nvidia GeForce RTX 3090 GPU in the TensorFlow framework (version 2.4.1) with Adam optimizer \cite{kingma2014adam} for 200 epochs using mean squared error loss function, a batch size of 32 and a learning rate of \num{5e-5}. The validation was performed on 15\% of the training data. The number of feature maps of the latent space was optimized using Keras-tuner \cite{chollet2015keras} and the checkpoint with the minimal validation loss was finally selected for inference.

\textbf{Raw signal networks -}
While \emph{b0-net} was trained using b0 images, \emph{Avg-b1000-net} was trained on average b1000 images, as training directly on individual b1000 images did not consistently converge \cite{kebiri2021through}. We have thus trained \emph{Avg-b1000-net} on average b1000 images with the aim of increasing the SNR and reducing variability. The average was computed over $n$ randomly selected volumes, $n\in \{3,6,15,30,40\}$. Empirically, higher $n$ means a lower risk of network divergence, at the cost of increased smoothness/risk of losing image detail. Therefore $n$ must be tuned. In the end, \emph{b0-net} was used to infer b0 images whereas \emph{Avg-b1000-net} was used to infer b1000 volumes.

\textbf{Spherical harmonics \index{spherical harmonics} network -}
We have fit SH \index{spherical harmonics} representations by using $L_{max}$=4 to the dMRI signal using Dipy \cite{garyfallidis2014dipy} and fed the resulting 15 SH \index{spherical harmonics} coefficients, slice by slice, to \emph{SH4-net}. Let us note that we preliminary computed the mean squared error difference with respect to the ground truth data when estimating SH\index{spherical harmonics} and projecting back to original grid from SH\index{spherical harmonics} bases of $L_{max} \in \{4, 6, 8\}$. As differences were relatively low between them (9.80, 8.64 and 9.95 for $L_{max} \in \{4, 6,8\}$ respectively, scale $\times10^{-4}$) and we aim at further testing on fetal data (where only 15 DWI are available) we selected to stick in what follows to $L_{max}$=4.

\begin{figure}[h!]
  \centering
    \includegraphics[width=1\columnwidth,keepaspectratio]{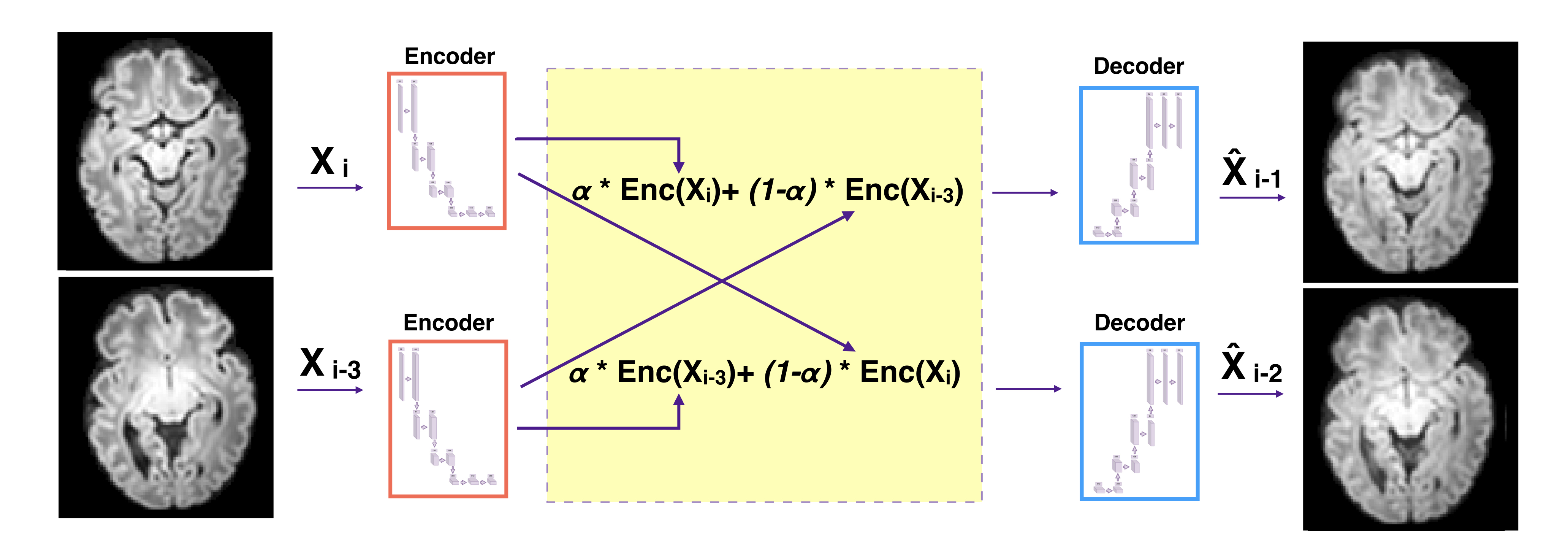} 
  \caption{Inference for two adjacent slices of the first coefficient of SH-$L_{max}$ order 4 illustrated for the case of $N=2$ where $\alpha=\frac{2}{3}$.}
  \label{fig:inference}
\end{figure}

\textbf{Inference in neonates -} For all networks (\emph{b0-net}, \emph{Avg-b1000-net} and \emph{SH4-net}), nested cross validation was performed where the 31 subjects were split into 8 folds. For each subject and each volume in the testing set, we removed N intermediate slices, $N \in \{1,2\}$ that were considered as the ground truth we aim to predict. Using the two adjacent slices, we input each separately to the encoder part of the network to get the $M$ latent feature maps. These feature maps were averaged using an equal weighting for $N=1$ and a $\{\frac{1}{3},\frac{2}{3}\}$, $\{\frac{2}{3},\frac{1}{3}\}$ weighting for $N=2$  (Figure \ref{fig:inference}). The missing slices were then recovered by using the decoder part from the resulting latent feature maps. The output of the network was then mapped back to the range of input intensities. This was performed using histogram matching (using cumulative probability distributions) between the network output as a source image and the (weighted) average of the two adjacent input slices as a reference image. Finally, the histogram matched output of \emph{SH4-net} was projected back to the original grid of 88 directions to recover the dMRI signal in the image domain.

\textbf{Evaluation in neonates -}
The inferred slices of \emph{Avg-b1000-net} were compared to conventional interpolations\index{interpolation}, namely trilinear, tricubic and B-spline of $5^{th}$ order \cite{tournier2012mrtrix,avants2009advanced}. The comparison was performed separately for one and two missing slices ($N \in \{1,2\}$) using the mean squared error (MSE). As all interpolation\index{interpolation} baselines produce similar results with a slight overperformance for the linear method (for $N=2$, MSE of 0.003164, 0.003204 and 003211 for linear, cubic and B-spline respectively), the former was chosen for further comparison with autoencoders.
 The two networks were additionally compared for FA\index{fractional anisotropy} and MD\index{mean diffusivity} maps that were extracted from the diffusion tensors \index{diffusion tensor imaging}, as estimated in Dipy \cite{garyfallidis2014dipy}. The DTI fit used the synthesized b0 by \emph{b0-net}. The linear baseline was further compared with \emph{SH4-net} and with the signal recovered from the same interpolation\index{interpolation} of the SH\index{spherical harmonics} coefficients. The comparison was also extended for DTI maps (FA\index{fractional anisotropy}, MD\index{mean diffusivity}). To compute them, DTI fit of \emph{SH4-net} relied on the b0 as synthesized by \emph{b0-net}, and the linear SH4 used corresponding linear interpolated\index{interpolation} b0. All comparisons were done using MSE for FA\index{fractional anisotropy} and MD\index{mean diffusivity} maps in white matter, cortical gray matter, and corpus callosum. Moreover, we have fit SH\index{spherical harmonics} representations of the ground truth signal by using $L_{max} = 4$ which were compared after projecting back to the original grid of 88 gradient unit vectors to the original DWI signal, separately for ($N \in \{1,2\}$). This was considered as the lower bound error of \emph{SH4-net}.


\textbf{Application to fetal DWI - } After fitting the SH\index{spherical harmonics} coefficients with $L_{max}$=4 to the fetal data. We have used \emph{SH4-net}, i.e., trained on pre-term neonates to infer SH\index{spherical harmonics} coefficients of middle ($N \in \{1,2\}$) slices of fetal subjects. The inference was performed in a similar manner as for neonates (Figure \ref{fig:inference}). Cropping of fetal images to $128\times128$ voxels was necessary before feeding them to the encoder. Then, we generated the diffusion tensor\index{diffusion tensor imaging} based on this new DWI signal and b0 using \emph{b0-net}, and visually assessed the consistency of the new slices in MD\index{mean diffusivity} and FA\index{fractional anisotropy} maps for the four subjects. Only qualitative evaluation was performed for fetal enhancement because of the lack of ground truth. 

\section{Results}

Based on the validation loss, the optimal number of feature maps in the latent space was found to be 32 for \emph{b0-net} and \emph{Avg-b1000-net}, and 64 for \emph{SH4-net}. For \emph{Avg-b1000-net}, averaging $n=15$ DWI was also found to be optimal. Moreover, the transposed convolution in the decoder did not reduce the validation loss as compared to performing a nearest neighbor interpolation\index{interpolation}. Hence all networks used the latter in the decoder part to avoid unnecessary overparameterization of the network. 

\subsection{DWI assessment}\label{sec:dwi_results}
\begin{figure}[]
  \centering
    \includegraphics[height=6.60cm,keepaspectratio]{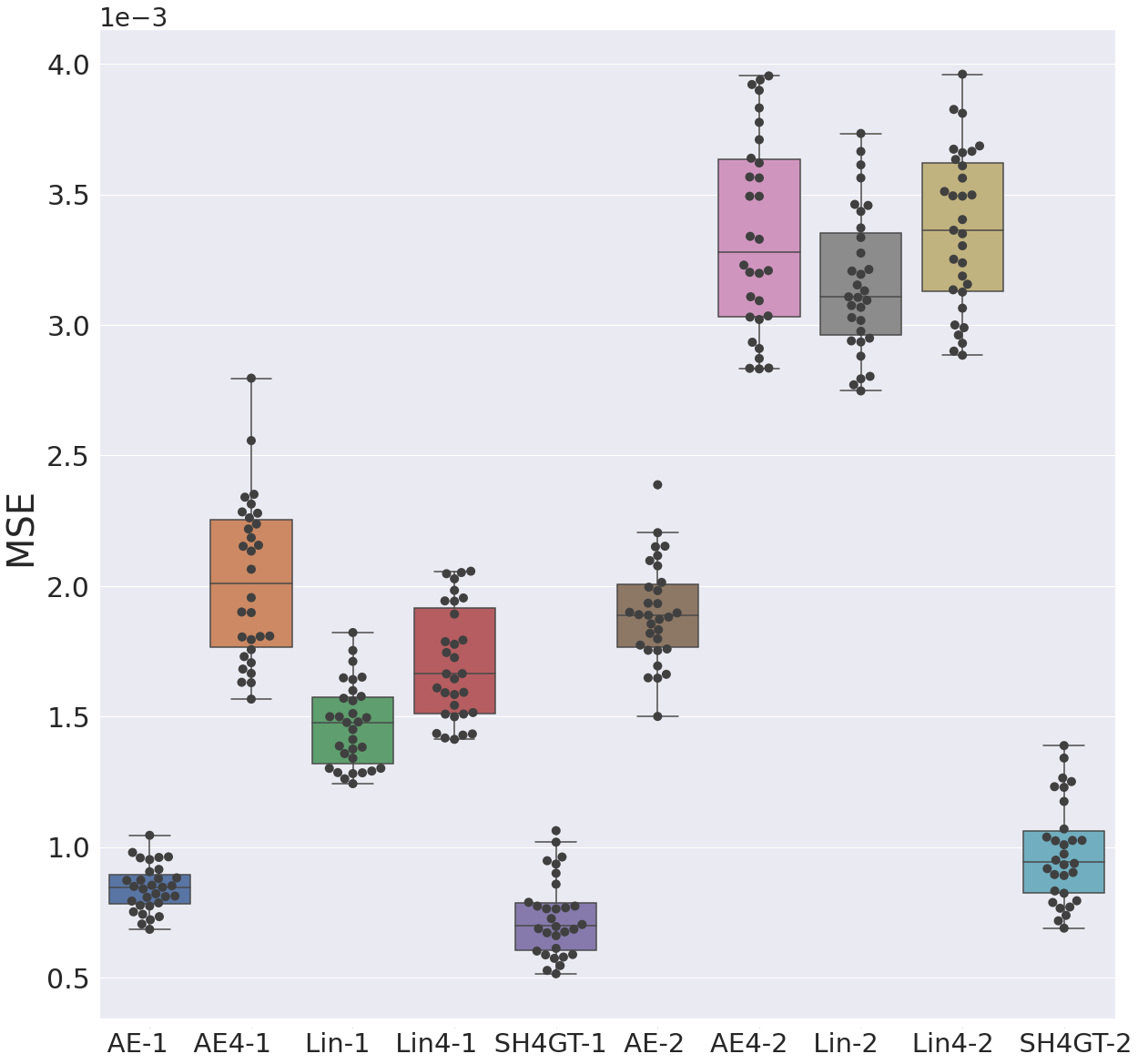} 
  \caption{Mean squared error (MSE) on dMRI images of autoencoder \index{autoencoder} enhanced using \emph{Avg-b1000-net} slices (AE-1, AE-2 for $N=1,2$ respectively) and for the baseline interpolation\index{interpolation} (linear on raw signal: Lin-1, Lin-2) and for \emph{SH4-net} and SH\index{spherical harmonics} linearly interpolated \index{interpolation} (Lin4-1, Lin4-2 for $N=1,2$ respectively). The lower bounds for the SH\index{spherical harmonics} errors (SH4GT) were also included as a reference. (Method-1, Method-2 for synthesizing/interpolating\index{interpolation} $N=1$ and $N=2$ slices, respectively)}
  \label{fig:raw_results}
\end{figure}

 Autoencoder\index{autoencoder} average b1000 trained network (\emph{Avg-b1000-net}) produces superior results compared to linear interpolation\index{interpolation} (Figure \ref{fig:raw_results}). The difference is higher for the case of two slices removed ($N=2$).

Comparing raw and SH\index{spherical harmonics} domain enhancement (Figure \ref{fig:raw_results}), we first observe that independently of the method (autoencoder\index{autoencoder} or linear), working directly on the raw signal outperforms working on SH\index{spherical harmonics} and projecting back to signal. In fact, autoencoder\index{autoencoder} \emph{Avg-b1000-net} outperforms linear interpolation\index{interpolation}, and for $N=1$ it is closely comparable to the SH\index{spherical harmonics} encoding (SH4GT-1 in Figure \ref{fig:raw_results}). While the SH\index{spherical harmonics} autoencoder\index{autoencoder} enhancement underperforms the classical SH\index{spherical harmonics} linear interpolation\index{interpolation} for $N=1$, \emph{SH4-net} slightly outperforms linear-SH for $N=2$. This gap between $N=1$ and $N=2$ for SH\index{spherical harmonics} linear and autoencoder\index{autoencoder} can be explained by the rich information that the autoencoder\index{autoencoder} was exposed to in the training phase from similar images compared to the interpolation\index{interpolation} that has solely access to local information.

\subsection{FA \index{fractional anisotropy} and MD \index{mean diffusivity} in newborns}\label{sec:dti_results}
\begin{figure}[h!]
    \centering
    \includegraphics[width=1\columnwidth]{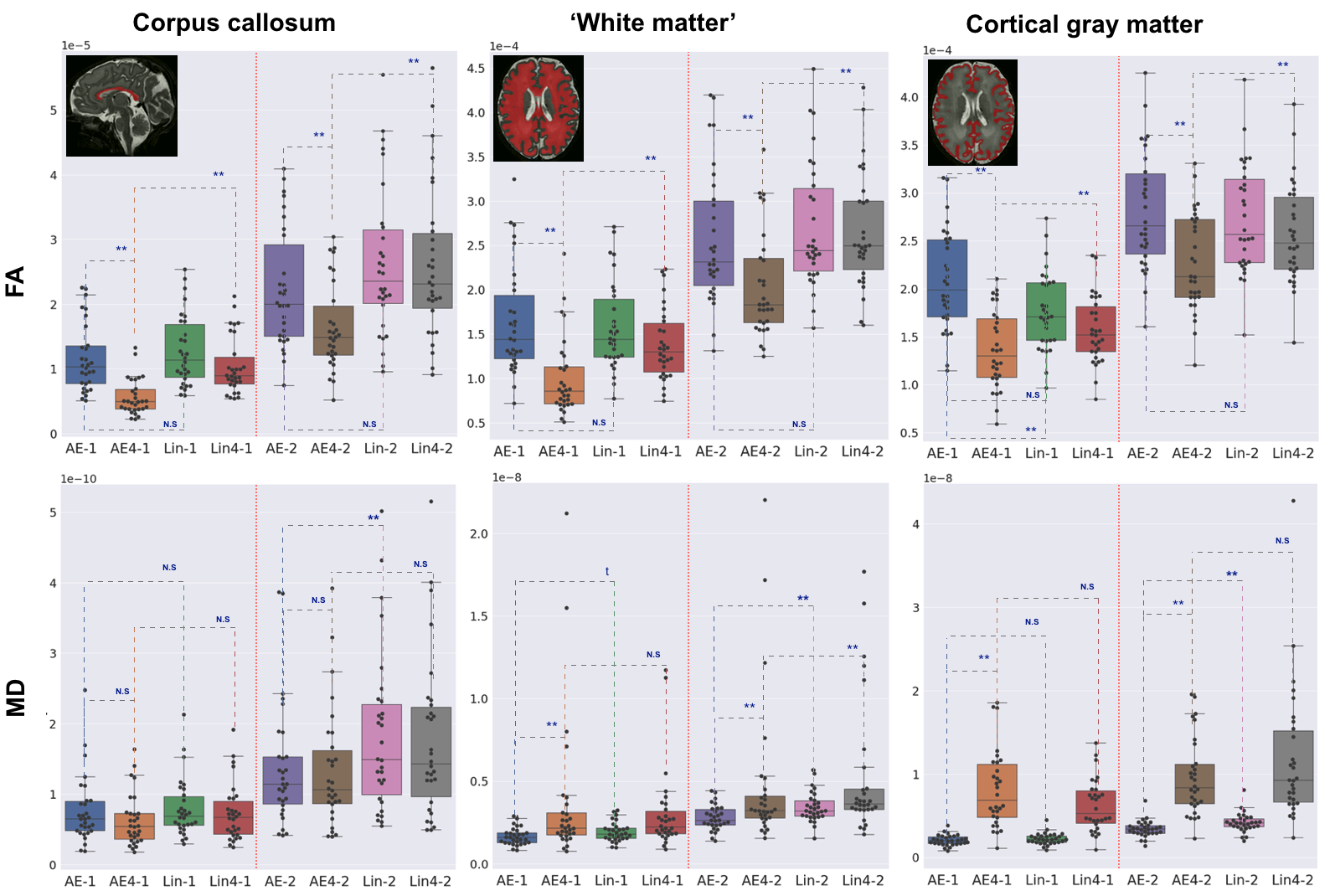}
    \caption{Mean squared error of fractional anisotropy\index{fractional anisotropy} (FA) and mean diffusivity\index{mean diffusivity} (MD) for different methods in three brain regions. See caption Figure \ref{fig:raw_results} for methods description. (Paired Wilcoxon signed-rank test: **: significant, p$<$0.028 - t: trending, p$=$0.06 - N.S.: non significant: p$>$0.06)}
 \label{fig:fa_md_results}
\end{figure}

Comparing DTI scalar maps (Figure \ref{fig:fa_md_results}) for the same previous configurations (see Figure \ref{fig:raw_results}), we notice that the autoencoder\index{autoencoder} enhancement outperforms the linear interpolation\index{interpolation} in all brain regions (except MD\index{mean diffusivity} for cortical gray matter when removing one slice, i.e. $N=1$) regardless of whether raw signal or SH\index{spherical harmonics} was used. This outperformance is significant (paired Wilcoxon signed-rank test) for FA\index{fractional anisotropy} in all SH \index{spherical harmonics} configurations, and for MD\index{mean diffusivity} in one third of all configurations. The difference is typically more pronounced when we remove two slices ($N=2$). Let us note that, opposite of what we observed at the DWI signal level, \emph{SH4-net} outperforms linearly interpolated\index{interpolation} SH\index{spherical harmonics}. Furthermore, for the FA\index{fractional anisotropy} map, \emph{SH4-net} obtains the lowest mean squared errors, thus it is more suitable than autoencoder\index{autoencoder} \emph{Avg-b1000-net} or the linear interpolation\index{interpolation}. The opposite trend, i.e. \emph{Avg-b1000-net} outperforming \emph{SH4-net} with statistical significance, can be noticed for MD\index{mean diffusivity}, with exception of the corpus callosum.

\subsection{Qualitative results of FA\index{fractional anisotropy} and MD\index{mean diffusivity} in fetuses}
\begin{figure}[h!]
    \centering
    \includegraphics[width=1\columnwidth]{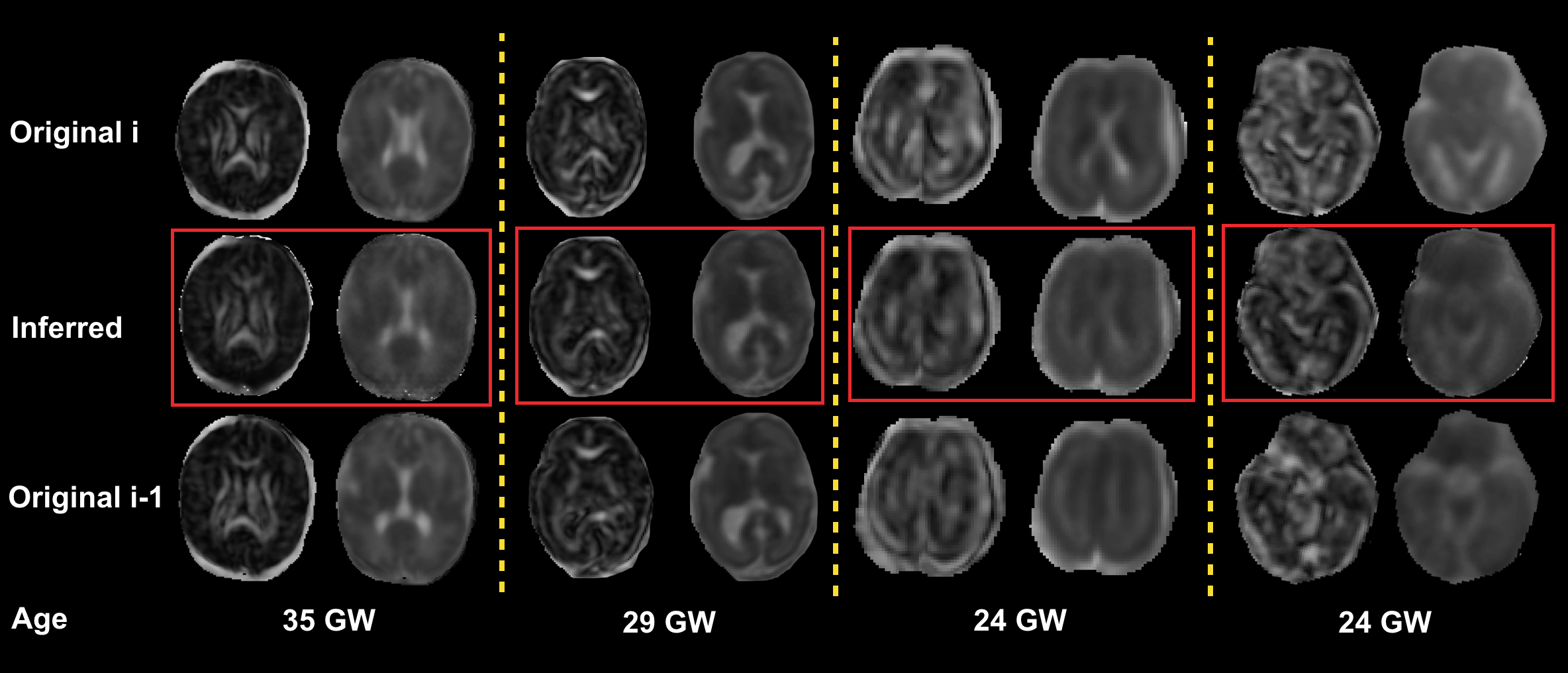}
    \caption{Fractional anisotropy \index{fractional anisotropy} (FA) and mean diffusivity\index{mean diffusivity} (MD) for four fetal subjects of respectively, from left to right, 4, 5, 4 and 4 mm of slice thickness. The middle row (red frames) illustrates synthesized slices corresponding to the diffusion tensor\index{diffusion tensor imaging} reconstructed with inferred DWI volumes with \emph{SH4-net} and b0 with \emph{b0-net}, using the two neighboring original slices (top and bottom rows).}
 \label{fig:fetal}
\end{figure}

The DWIs synthesized by \emph{SH4-net} using the latent space were visually consistent as they smoothly vary between the adjacent slices. Figure \ref{fig:fetal} displays the corresponding FA\index{fractional anisotropy} and MD\index{mean diffusivity} maps for four subjects. We can clearly delineate the smooth transition between the two adjacent slices, especially in late gestational weeks fetuses in which the structures are more visible. For instance, the corpus callosum and the internal capsules of the synthesized slices displayed in FA\index{fractional anisotropy} maps are coherent with respect to their neighbouring slices. 

\section{Conclusion}
We have proposed autoencoders\index{autoencoder} for dMRI through-plane slice inference in early brain development. The assessment was performed in both raw signal and spherical harmonics\index{spherical harmonics} (SH) domains, where the latter proved to be more accurate for DTI-FA\index{fractional anisotropy} maps reconstruction and the former for raw data estimation. We hypothesize that this could be explained by some global bias introduced to the back projected raw signal by the SH\index{spherical harmonics} trained autoencoder\index{autoencoder}. However, the orientation information (i.e., signal's shape) was better preserved and hence, FA\index{fractional anisotropy} which is scale invariant, was clearly better depicted by SH\index{spherical harmonics} autoencoder\index{autoencoder} estimation. Lastly, we have successfully applied our method trained on newborn data to enhance the through-plane resolution of fetal data acquired in a different scanner, with a lower b-value and fewer gradient directions. Inferring missing slices or realistically increasing the through-plane resolution has to potential to translate to more accurate diffusion properties and hence a better uncovering of the underlying brain structure. In future work, we aim to increase the angular resolution in fetal images by using \emph{supervised learning} to map spherical harmonics\index{spherical harmonics} coefficients of order 4 (i.e., the maximal order that can be fit with clinical fetal images) to higher orders (6 or 8) using pre-term data.



\section{Acknowledgments}

This work was supported by the Swiss National Science Foundation (project 205321-182602, grant No 185897: NCCR-SYNAPSY- "The synaptic bases of mental diseases" and the Ambizione grant PZ00P2\_185814). We acknowledge access to the facilities and expertise of the CIBM Center for Biomedical Imaging, a Swiss research center of excellence founded and supported by Lausanne University Hospital (CHUV), University of Lausanne (UNIL), École polytechnique fédérale de Lausanne (EPFL), University of Geneva (UNIGE), Geneva University Hospitals (HUG) and the Leenaards and Jeantet Foundations.
\clearpage
%
\bibliography{mybibliography.bib}{}
\bibliographystyle{splncs04}

\end{document}